\documentclass[preprint]{aastex}
\input epsf

\shorttitle{Revisiting Globular Clusters of NGC 3610}
\shortauthors{Strader et al.}
\def\etal{{\it et al.~}}

\begin{document}

\title{Revisiting the Globular Cluster System of the Merger Remnant Elliptical NGC 3610}

\author{Jay Strader and Jean P. Brodie}
\affil{UCO/Lick Observatory, University of California, Santa Cruz, CA 95064}
\email{strader@ucolick.org, brodie@ucolick.org}

\and

\author{Duncan A. Forbes}
\affil{Centre for Astrophysics and Supercomputing, Swinburne University, Hawthorn VIC 3122, Australia}
\email{dforbes@astro.swin.edu.au}

\begin{abstract}

We have obtained Keck spectra of six candidate globular clusters (GCs) in the dynamically young elliptical galaxy NGC 3610,
supplementing our previous Keck spectra of eight confirmed GCs (Strader \etal 2003). Five of our new candidates are confirmed
to be GCs. Of the thirteen GCs, eleven are located within the $K$ band effective radius of 7 kpc. Two of these thirteen
clusters are found to be young ($\sim 2$ Gyr) and very metal-rich ([$Z$/H]$\sim$ +0.5), three are old and metal-poor, and the
remaining eight clusters are old and metal-rich. The ages of the young clusters are consistent with a recent spectroscopic age
estimate of 1.6$\pm$0.5 Gyr for the galaxy itself (Denicolo \etal 2003) and suggest that these clusters formed in the
disk-disk merger which likely created NGC 3610. Intriguingly, \emph{both} young GCs have [$\alpha$/Fe] $\sim +0.3$, while the
majority of the old clusters are not $\alpha$-enhanced, in contrast to Galactic and M31 GCs, and contrary to predictions of
nucleosynthetic calculations. The two old subpopulations of GCs can be attributed to the merger progenitors. The relative
numbers of old and new metal-rich GCs are poorly constrained because of the expected differences in radial distributions of
the two subpopulations. However, based on our spectroscopic results and a comparison of the \emph{Hubble Space Telescope}
color magnitude diagram (Whitmore \etal 2002) with stellar population models, we argue that more than half of the metal-rich
GCs are likely to be old.

\end{abstract}


\section{Introduction}

Schweizer (1987) and Ashman \& Zepf (1992) suggested that the formation of globular clusters (GCs) in gaseous disk-disk galaxy mergers
could explain several puzzles about GC systems, for example, the high specific frequencies of GCs in ellipticals. The discovery of young
massive stellar clusters in ongoing disk-disk mergers (e.g., the Antennae; Whitmore \& Schweizer~1995) and bimodal GC color distributions
in early-type galaxies is often seen as supporting this merger scenario for the formation of metal-rich GCs. However, substantial
observational and theoretical questions regarding this scenario remain. Are GCs in mergers formed with the efficiencies needed to
overcome the specific frequency problem? What proportion of newly born clusters will survive destruction by evaporation and tidal shocks?
How can the substantial similarities between the GC color distributions of spirals and ellipticals be explained?

After just a few Gyrs, a disk-disk merger is thought to resemble an elliptical galaxy, both morphologically and dynamically (Barnes \& Hernquist
1992). Merger remnant ellipticals with ages from $\sim 1 - 5$ Gyr provide a key consistency check of the ``formation by merger''
scenario. Such galaxies are young enough that the observational signs of the interaction (e.g., morphological fine structure, evidence of a
fading starburst) have not disappeared but are old enough that substantial dynamical evolution of the GC system has taken place.

NGC 3610 is a prime example of such a system. Based upon its disturbed morphology, enhanced Balmer line strengths, and bluer-than-average $UBV$ colors, Schweizer \&
Seitzer (1992) suggested that it was a disk-disk merger remnant, with an estimated merger age of $4\pm2.5$ Gyr. $JHK$ imaging (Silva \& Bothun 1998) suggests the
presence of an central intermediate-age population. Using Lick/IDS absorption line indices and the single stellar population (SSP) models of Thomas, Maraston \& Bender
(2003), Denicolo \etal (2003) find a luminosity-weighted age of $1.6\pm0.5$ Gyr (random errors only) within a standard r$_{e}/8$ central aperture, though systematic
errors due to the assumption of a single starburst and in the SSP models themselves could be substantial. Whitmore \etal (1997)  and Whitmore \etal (2002, hereafter W02)
presented \emph{Hubble Space Telescope} (HST) observations of the galaxy, in which they discovered a population of luminous red objects concentrated toward the center of
the galaxy. The luminosity function (LF) of these objects appears to follow a power law---similar to the young GCs in the Antennae but unlike the log-normal form of the
GCLF seen in most galaxies. Based on these findings, they argued that these red objects might be metal-rich GCs formed in the merger.

To test this proposed scenario, we previously obtained spectra of a sample of eight luminous GCs chosen from the HST studies (Strader \etal 2003,
hereafter Paper I). Six of these objects turned out to be metal-rich but, perhaps surprisingly, only one of these was found to be young ($\sim 2$
Gyr) and likely formed in the merger. The other five metal-rich GCs appear old, with ages comparable to metal-rich GCs in the Galaxy. At first
glance, this finding suggests that many of the red GCs in the HST images may be old metal-rich GCs (presumably from the progenitor spirals) and
that few new GCs formed in the merger. The young cluster also appears to have a supersolar [Mg/Fe] abundance ratio. However, due to the
observational difficulties inherent in obtaining spectra of objects which are extremely faint ($V \sim 22-24$) and located in areas of high
background, most of the GCs selected for spectroscopic study were located some distance from the galaxy center, i.e., between 22\arcsec and
89\arcsec. The galaxy effective radius (r$_{e}$) is $\sim 13 \arcsec$ (1.9 kpc) in the $B$ band (Faber \etal 1989). This gives a measure of the
size of the central star-forming region, as the $B$ band light will be dominated by younger stars associated with the intermediate-age starburst.
The $K$ band effective radius for NGC 3610 is 66\arcsec or 9.6 kpc (Brown \etal 2003), which gives a better measure of the size of the underlying
mass distribution. It is difficult to estimate how well we are sampling the intermediate-age population of GCs; while simulations of disk-disk
mergers suggest that $\sim 90\%$ of the metal-rich GCs created in the merger lie within 3-4 kpc of the galaxy center (Bekki\etal 2003), the young
GCs in the Antennae are distributed throughout the $\sim 7.5 \times 7.5$ kpc WFPC2 field of view (Zhang, Fall \& Whitmore 2001). This suggests
that we might expect young GCs in NGC 3610 to be found at a variety of radii, indeed, those further from the center may be better able to survive
the violent dynamical destruction processes during and after the merger.

To further probe the age distribution of the GCs in NGC 3610, we have obtained a second sample of six GC candidates in the inner regions of the
galaxy. In the present paper, we determine metallicities, ages, and abundance ratios for this new sample of GCs, and discuss the implications for
GC formation in mergers.

\section{Observations and Data Reduction}

The six GC candidates were observed with the Keck I telescope and the Low Resolution Imaging Spectrometer (LRIS, Oke \etal 1995) during a three
night run in 2003 January. These objects were chosen from the list of GC candidates identified by W02 on dithered HST Wide Field Planetary
Camera~2 (WFPC2) images, and were observed through a slitmask with $0.8 \arcsec$ slits. Seeing ranged from $0.7 \arcsec - 0.8 \arcsec$ during
the run.

For each of the six candidates, nineteen 30-minute exposures (totaling 570 minutes) were taken. Observations were made with a 600 line mm$^{-1}$
grism blazed at 4000 \AA\ on the blue side and an 831 line mm$^{-1}$ grating blazed at 8200 \AA\ on the red side, with a dichroic splitting the
incoming beam at 6800 \AA. The resulting wavelength ranges were approximately 3300--5900 \AA\ and 7100--9100 \AA, respectively, though there were
often variations of several hundred \AA\ due to the differences in the location of individual slitlets within the mask. The red side wavelength
range was chosen to cover the \ion{Ca}{2} triplet (we defer the analysis of the red side spectra to a future paper). The spectral resolution of
the blue spectra was 3.3 \AA.

The standard data reduction was performed using image and spectral reduction tools in IRAF.\footnote{IRAF is distributed by the National Optical
Astronomy Observatories, which are operated by the Association of Universities for Research in Astronomy, Inc., under cooperative agreement with
the National Science Foundation.} Raw images were debiased and then flatfielded, using normalized composite sky and dome flats. While the
majority of the spectra were successfully extracted, for a few of the faintest objects ($V \sim 23-23.5$) the continuum could not be traced in
several exposures taken during the final night of the run. Due to difficulties in finding a suitable reference spectrum for tracing, we coadded
several pairs of consecutive exposures and extracted the spectra from these combined images. We verified the stability of the wavelength solution
within each pair by comparison of the wavelength calibrated spectra of the two brightest object spectra.

The extracted spectra were wavelength calibrated using spectra of HgNeArCdZn comparison arc lamps (essentially all lines in the blue spectra were
Hg, Cd, and Zn). After this wavelength calibration, small ($1-2$ \AA) zero point corrections in the science spectra were made using the bright
\ion{O}{1} skyline at 5577.34 \AA. The spectra were then coadded with sigma clipping. The coadded spectra were flux-calibrated using observations
of the flux standard Feige 34, chosen from Massey \& Gronwall (1990) as a standard calibrated up to 1 $\mu$m. To assess the quality of the 
observations, the signal-to-noise ratio (S/N) of the final coadded spectra was measured in the range $4700-5300$ \AA\ (for consistency with Paper 
I, which had a redder wavelength range).

\section{Radial Velocities and \ion{H}{1}}

The radial velocities of individual clusters were measured by cross-correlating the cluster spectra with spectra of two radial velocity standards
and confirmed by cross-correlating the cluster spectra with each other. The final heliocentric radial velocity of each individual GC (listed in
Table 1) is a weighted average of the two measurements.

The spectrum of W50 was very low S/N, and no significant cross-correlation peak could be found and thus no radial velocity measured.  
Qualitatively, its spectral shape was similar to that of an old stellar population, but no familiar absorption lines were readily identifiable.
It may be a foreground star or a compact background galaxy. Excluding this object, five of our candidates appear to be \emph{bona fide} GCs.

Table 1 lists basic data for the five confirmed GCs, as well as (for comparison purposes) the data for GCs from Paper I. The positional and
photometric data are taken from W02, while the heliocentric radial velocities are based on our own measurements, as described in this section. No
extinction corrections were applied since \citet{BH84} give $A_{V}$ = 0.00 mag and \citet{S98} give $A_{V}$ = 0.03 mag. Spectra of clusters W11, 
W12, and and W22 are presented in Figure 1; spectra of three clusters from the first GC sample (W6, W9, and W10) are given in Figure 1 of Paper 
I.

Combining these five new GCs with our Paper I sample gives thirteen clusters with measured radial velocities. The mean heliocentric radial velocity
of these clusters is $1754\pm20$ km s$^{-1}$, with a velocity dispersion of $69\pm27$ km s$^{-1}$ (compared to values $1769\pm15$ km s$^{-1}$ and
$45\pm25$ km s$^{-1}$ found in Paper I). This new mean velocity compares reasonably well to the systemic velocity of the galaxy, $1696\pm17$ km
s$^{-1}$, listed in the Updated Zwicky Catalogue \citep{F99}.

Interestingly for the proposed disk-disk merger scenario for the birth of NGC 3610, the galaxy seems to currently contain very little \ion{H}{1}.
Hibbard \& Sansom (2003) found no evidence of \ion{H}{1} emission in VLA observations, giving an upper mass limit of $2 \times 10^{7}$
M$_{\odot}$ of neutral gas. If the progenitor spirals had typical Sb/Sc \ion{H}{1} gas masses of a few times $10^{9}$ M$_{\odot}$, over 99\% of
the incoming gas must have been turned into stars, ionized, or removed from the galaxy. Using near-IR photometry, Silva \& Bothun (1998) were
able to detect an intermediate-age population of stars, but could only put weak limits on the total size of the starburst ($\la 6 \%$ of the
total stellar mass). The low X-ray luminosity of the galaxy (Fabbiano \& Schweizer 1995) implies that any ionized gas is not in a hot gaseous
halo. Therefore, if NGC 3610 was indeed created by the merger of two spirals, $\ga 10^{9}$ M$_{\odot}$ of gas must have been removed, possibly
blown out of the potential well of the galaxy by the merger-induced starburst. Alternatively, if one (or both) of the progenitors was an S0
galaxy, less incoming gas would be expected, although the existence of an intermediate-age stellar population suggests that some gas must have
been involved.

\section{Cluster Metallicities, Ages, and Abundances}

\subsection{Brodie \& Huchra Metallicities}

\citet[hereafter BH90]{BH90} defined an empirical procedure for measuring the metallicity of globular clusters by taking a weighted mean of
various elemental absorption-line indices sensitive to overall metallicity, such as Ca H+K, CN, and Mg. Indices defined over a wavelength range
of interest are calculated with respect to a pseudocontinuum, defined using regions to either side of the feature passband. Since BH90 calibrated
their metallicities to the Zinn \& West (1984) scale, BH90 [Fe/H] values are more likely a measure of ``composite'' metallicity than pure Fe
abundance, a fact noted in their paper. Since BH90 primarily used clusters with [Fe/H] $\la -0.5$ to calibrate the relation (an exception is the
solar-metallicity old open cluster NGC 188), metallicity estimates for GCs more metal-rich than this are considerably uncertain and tend to
overestimate the cluster's true metallicity. [Fe/H] measurements for young or intermediate-age GCs are uncalibrated and thus may suffer from
unknown systematic uncertainties. However, for old metal-poor GCs, the BH method appears to accurately measure [Fe/H] on the Zinn \& West scale
(see, e.g., Figure 6 in Thomas \etal 2003). Indeed, for this class of GCs, the robustness of the BH90 procedure makes these metallicites
generally preferable to those obtained using Lick/IDS index plots (see below).

No resolution corrections were made for measuring BH90 metallicity indices on our GCs, since the system was calibrated with data of 2--5 \AA\
resolution, similar to that of our spectra. After zero-redshifting the clusters, we measured [Fe/H] for our five clusters according to the
formulae in BH90, with two small changes. First, our calculation of the variance was modified slightly (see Larsen \& Brodie 2002). Second, for
several clusters, we excluded feature indices affected by poor subtraction of night sky lines or insufficient spectral coverage in the blue. The
resulting individual index [Fe/H] estimates, along with the composite [Fe/H] values, are given in Table 2. Since we later determine W11 to be a
young GC (see next section), we have excluded it from the table. The composite metallicities listed in Table 2 suggest that GCs W12 and W22 are
metal-poor and W33 and W40 are metal-rich.

\subsection{Lick/IDS Metallicities and Ages}

We have also measured Lick/IDS absorption-line indices (Worthey \etal 1994; Trager \etal 1998) which can be used to infer cluster metallicities,
ages and abundance ratios. Since Lick/IDS spectra were of substantially lower resolution (8--12 \AA) than our data (3.3 \AA), we smoothed our
fluxed spectra with a wavelength-dependent Gaussian kernel using the prescription of Worthey \& Ottaviani (1997) to more closely match the
Lick/IDS resolution. The measured Lick/IDS indices for the five new GCs are listed in Table 3. To verify our match to the Lick/IDS system, we
measured indices from spectra of several Lick/IDS standard stars taken during the run, and found the offsets to be less than the measurement
uncertainties listed in Table 3.

The metallicities and ages of the GCs in our sample are estimated using the variable elemental abundance SSP models of Thomas \etal (2003) in
order to match the supersolar [$\alpha$/Fe] ratios typical among GCs. The effects of abundance variations on the Lick/IDS indices were calculated
by the method described in Trager \etal (2000), based upon the spectral modeling work of Tripicco \& Bell (1995).

Figure 2 shows H$\beta$ plotted against $<$Fe$>$, an average of the indices Fe5270 and Fe5335, for twelve clusters. We have excluded the red GC
W28 from Paper I in this and subsequent line index plots since an accurate Fe5335 index could not be measured (however, based on its H$\beta$ and
Fe5270 indices, it appears to be old and metal-rich, not metal-poor as claimed in Paper I). Superimposed on this plot are isochrones and
isometallicity lines from Thomas \etal (2003), assuming [$\alpha$/Fe] = +0.3. The GC sample from Paper I divides cleanly into three groups: one
old metal-poor cluster, five old metal-rich clusters, and one young metal-rich cluster. For our new sample, two of the five clusters appear to be
old and metal-poor, and one (W11) is quite young (1--3 Gyr) and metal-rich. The other two clusters (W33 and W40) lie to the right of the grids,
suggesting that they are very metal-rich, however, this could be caused by anomalous abundance ratios (see below). The positions of GCs in the
H$\beta$ vs. $<$Fe$>$ diagram are dependent on the [$\alpha$/Fe] ratio, so variations in this ratio from cluster to cluster will affect the
age/metallicity determination.

Thomas \etal (2003) have defined a combined Mg+Fe index, [MgFe]$\arcmin$, which is largely independent of [$\alpha$/Fe] (it is very similar to
the [MgFe] index introduced in Gonz{\' a}lez 1993). In Figure 3 we plot H$\beta$ vs. [MgFe]$\arcmin$ for the total GC sample. Due to poor night
sky line subtraction, we were unable to measure Mg$b$ for GC W12, so it is not plotted (based on the the Brodie \& Huchra indices and Figure 2,
W12 is likely an old metal-poor GC). Both of the potentially very metal-rich clusters (W33 and W40) now have much lower inferred metallicities,
which are more consistent with their values from the Brodie \& Huchra method (see Table 2).

Generally, Figure 3 supports the results of Figure 2, i.e., NGC 3610 contains one young and two old subpopulations of GCs. The old subpopulations
are presumably from the progenitor galaxies that merged to form NGC 3610. The ages of the two young GCs, W6 and W11, are estimated to be 1--2 Gyr
and 1--3 Gyr old, respectively, with [$Z$/H] $\sim +0.4$ and +0.7. Denicolo \etal (2003) recently derived a Lick-based spectral age of
1.6$\pm$0.5 Gyr for the central young stellar population. These stars were estimated to have [$Z$/H] = +0.6$\pm$0.4. We have plotted her
r$_{e}/8$ central aperture point in Figures 2 \& 3, and it is consistent in both cases with the index measurements for the two young GCs. Thus,
within the errors, it appears reasonable to conclude that these few Gyr old, very metal-rich ([$Z$/H] $\sim$ +0.5) GCs indeed formed in the
gaseous merger that created NGC 3610.

For reference, a schematic diagram of the positions of GCs in our sample with respect to the galaxy center is shown in Figure 4. Overplotted are
the $B$ and $K$ band effective radii of NGC 3610, tracing the galaxy isophotes. Our combined sample of GCs has eleven out of thirteen within this 
latter radius, and one at the edge of the $B$ band r$_{e}$.

\subsection{[$\alpha$/Fe] and Formation Timescales}

To estimate [$\alpha$/Fe] for our GCs, we plot an Fe index ($<$Fe$>$) vs. an $\alpha$-element index (Mg$b$) in Figure 5, overlaid with Thomas \etal
(2003) models. Of the nine old GCs with [$\alpha$/Fe] measurements, only two appear to have the $\alpha$-enhancement typical of Galactic GCs.  The
remainder appear to have solar or subsolar abundances, although based upon Figures 2 \& 3, GCs W33 and W40 seem to have anomalously high $<$Fe$>$
values. Since these clusters also have the lowest S/N in our sample, we attach little significance to these possibly anomalous results.

Intriguingly, \emph{both} young GCs have [$\alpha$/Fe] $\sim +0.3$. If the ISM of the merger progenitors was enriched to $\sim$ solar [Fe/H] and
began with [$\alpha$/Fe] $\sim 0.0$, an enormous mass of $\alpha$-elements, explosively produced in Type II supernovae in the merger-induced
starburst, would be required to raise it back to [$\alpha$/Fe] $\sim +0.3$ before the GCs were formed (especially at these high metallicities).
Galactic nucleosynthetic calculations show that these supersolar abundance ratios are difficult but not impossible to achieve (Chiappini \etal 1999).
We note that the timescale of Fe production in Type Ia SNe is often taken to be $\la 1$ Gyr. While it may take this long for SNe Ia to first be
produced in appreciable quantities, the peak of SN Ia production does not occur until several Gyr later (see, e.g., Timmes, Woosley, \& Weaver 1995)
and thus the ``window'' for $\alpha$-enhanced GC production may be a few Gyr long after the initial starburst, making [$\alpha$/Fe] less useful as a
chronometer.

Are there other possible explanations for the observed [$\alpha$/Fe] enhancements in the young clusters? A top-heavy IMF in a generation of stars
preceding GC formation would produce supersolar [$\alpha$/Fe] values; tentative evidence for such IMF variations in young \emph{clusters} has been
found by, e.g., Brodie \etal (1998) and Smith \& Gallagher (2001). However, further evidence would be required before seriously considering this 
scenario.

It may also be possible that the suspected young clusters are not in fact young. Our age-dating method could underestimate the ages of old metal-rich
GCs with blue horizontal branches, like the Galactic bulge GCs NGC 6441 and NGC 6388. However, the H$\beta$ enhancement for these two Galactic GCs is
a relatively modest effect, and they would not look young enough on SSP model grids to be mistaken for very young (1--2 Gyr) GCs. For example, Thomas
\etal (2003) models give an age of 6 Gyr for NGC 6441 (R.~Schiavon, private communication), while its color-magnitude diagram suggests it is at least
as old as 47 Tuc (Rich \etal 1997). However, given our poor current understanding of the second parameter problem, it is impossible to rule out the
blue horizontal branch scenario.

Finally, we note that systematic effects in the Thomas \etal (2003) models could be responsible, at least in part, for the supersolar [$\alpha$/Fe]
values in the young GCs. The pentagon in Figure 5 represents the solar-metallicity, 3.5 Gyr old Galactic open cluster M67 (Schiavon, Caldwell, \& Rose
2003). M67 is known to have [$\alpha$/Fe] $\sim 0.0$ (Shetrone \& Sandquist 2000; Tautvai{\v s}iene \etal 2000), however, M67 is 3$\sigma$ deviant
from the solar, 4 Gyr [$\alpha$/Fe] isochrone in Figure 5 in the supersolar sense. Since Thomas \etal (2003) use the Tripicco \& Bell (1995) method
in modeling index changes due to variations in log $g$ and $T_{\rm{eff}}$, and do so only for a single isochrone (5 Gyr, solar metallicity), they
must assume that the index responses to stellar parameter variations are independent of age and metallicity. They argue that this is likely to be the
case for SSPs with ages $\ga 1-2$ Gyr (after the RGB phase transition), but additional theoretical and observational work is needed to confirm this
assertion, and extend the results to the supersolar metallicities suspected for our young GCs.

\section{Implications and Concluding Remarks}

Based on the ages and metallicities of the young GCs determined above, we can estimate the colors and luminosities expected for the GC
subpopulation presumed to have formed in the merger. While W02 performed a similar exercise, we now have more accurate,
spectroscopically-determined physical parameters. Noting that the mean values for the two GCs agree well with the luminosity-weighted central
galaxy parameters of Denicolo \etal 2003 ([$Z$/H] $\sim +0.6$, age $\sim 1.6$ Gyr), we adopt these as the simulation values. Next, using the SSP
models of Bruzual \& Charlot (2001) and adopting a typical $V-I$ photometric error for GCs with $V \le 25.5$ from W02, we can define a 1$\sigma$
region in $V$ and $V-I$ space in which we would expect the young GCs to fall. Figure 6 shows this region around $V-I = 1.20$ superimposed upon
the brighter part of the W02 color-magnitude diagram for NGC 3610 GCs. We also show the region defining a similar 1$\sigma$ distribution for old
metal-rich GCs, noting that a 13 Gyr, [$Z$/H] = $-$0.7 Bruzual \& Charlot model gives $V-I = 1.09$. This is consistent with the mean color
observed for Galactic metal-rich GCs ($V-I$ = 1.06; Barmby \etal 2000). The dashed line denotes the color cut adopted by W02 for the separation
of blue and red GCs, and the horizontal dot-dashed line shows the brightness $\omega$ Cen would have at the distance of NGC 3610.

While W02 assumed that all of the red GCs were young, our results show a substantial population of \emph{old} metal-rich GCs, presumably
from the progenitor galaxies of NGC 3610. Figure 6 shows that this old subpopulation mostly occupies the same area of color space as
Galactic metal-rich GCs, supporting our conclusion. Interestingly, both of our young GCs fall within the shaded region expected for
young, metal-rich GCs. Note also that this region seems to trace the red GCs on the PC chip (which was aligned on the galaxy center).
This group of GCs appears to be $\sim 1$ magnitude more luminous than the old metal-rich GCs (traced by our spectroscopically-confirmed
members), consistent with Bruzual \& Charlot model predictions using the parameters listed above. Together, these observations suggest
that the young GCs in NGC 3610 are centered around $V-I = 1.20$.

We note that the Local Group Elliptical (i.e., the dissipationless merger of all Local Group GC systems), defined by Forbes \etal (2000),
has a ratio of old blue to old red GCs of 2.5:1. If the progenitors of NGC 3610 were intermediate Sb/Sc type spirals (like M31 and the
Galaxy) then we might expect a similar blue to red ratio of old GCs in NGC 3610. Comparing relative numbers of old and young red GCs is
difficult, given the expected differences in LFs and radial distributions of the two subpopulations. However, it is clear that only a
subset of the red GCs (as defined by W02) in NGC 3610 are in fact young. An examination of Figure 6 suggests that half or more of the
GCs with $V-I > 1.025$ may in fact be old, so the number of young GCs claimed by W02 could be overestimated by at least a factor of two.

W02 found that the metal-rich GCLF was best fit by a power law with index $\alpha \sim -1.9$. Our results show that the red LF is in fact a
combination of an old metal-rich subpopulation (which likely has a Gaussian LF) and a young metal-rich subpopulation. Our observations need not
invalidate the W02 finding of an \emph{overall} power-law form of the red LF, since it is possible that the young GCs have a power-law form which
dominates the combined LF. Indeed, the best-fit Fall \& Zhang (2001) disruption model gives an age of 1.5 Gyr for the young GCs in NGC 3610,
consistent with our findings, though it appears that a lognormal initial GCLF is a better fit to observations of the giant elliptical M87
(Vesperini \etal~2003). However, it is not clear whether the slope of the metal-rich LF would change if the probable Gaussian form of the old
metal-rich GCLF was subtracted off. In future studies of suspected merger remnants, the possible presence of substantial numbers of old
metal-rich GCs, coupled with the age-metallicity degeneracy in optical colors, need to be considered carefully.

\acknowledgements

We acknowledge support by the National Science Foundation through Grant AST-0206139 to JPB. This material is based upon work supported under a
National Science Foundation Graduate Research Fellowship to JS. We acknowledge the helpful comments of an anonymous referee. We thank Soeren
Larsen for his aid in the design of the slitmask, and Francois Schweizer, Ricardo Schiavon, Michael Beasley, Michael Pierce, and Justin Howell
for useful comments on the manuscript. We also thank Scott Trager for pointing out the index measurement problem for cluster W28. Benjamin de
Bivort aided the creation of Figure 4. Data presented herein were obtained at the W.~M.~Keck Observatory, which is operated as a scientific
partnership among the California Institute of Technology, the University of California, and the National Aeronautics and Space Administration.
The Observatory was made possible by the generous financial support of the W.~M.~Keck Foundation.

\newpage

\epsfxsize=14cm   
\epsfbox{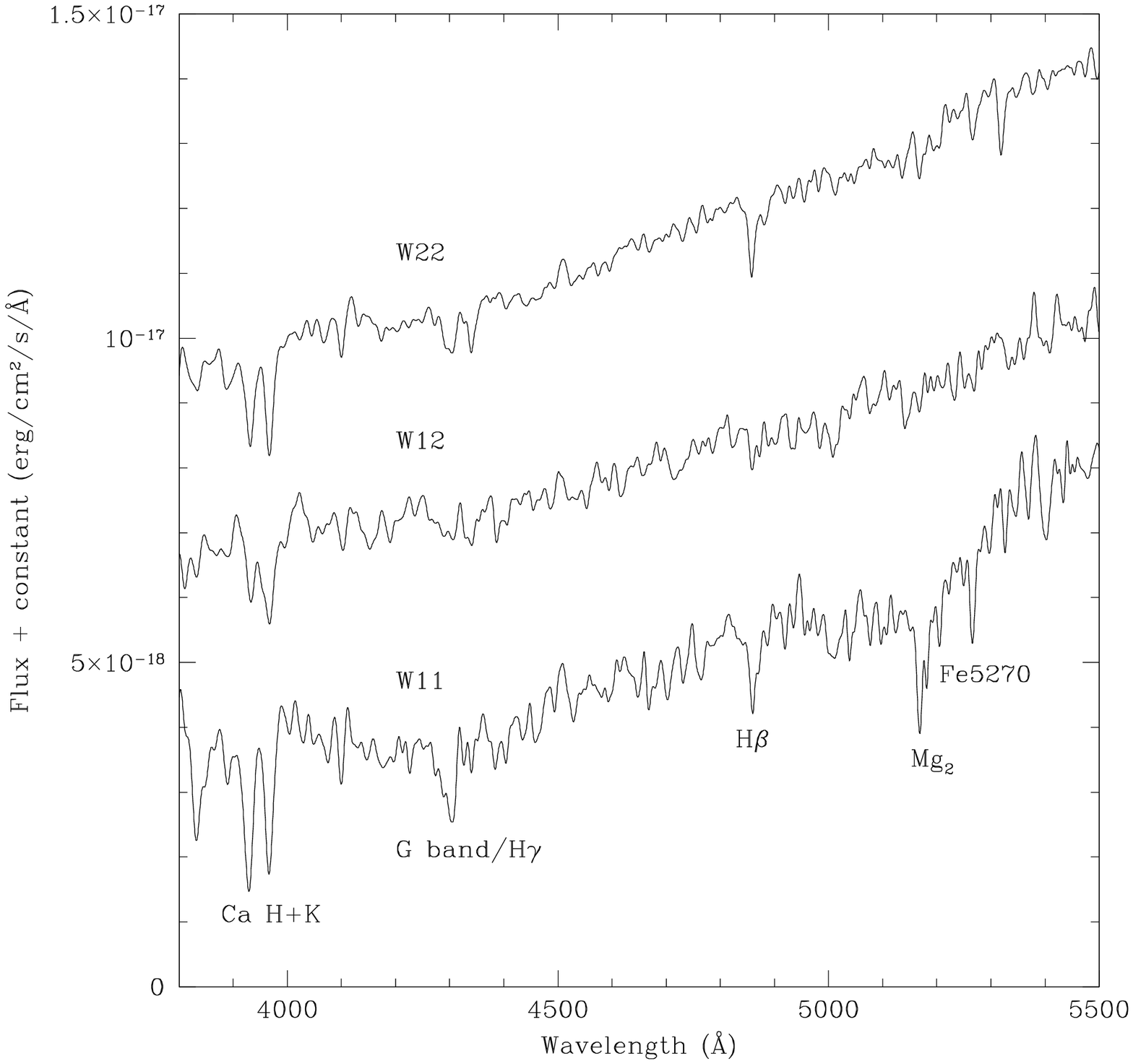}
\figcaption[strader.fig1.eps]{\label{fig:spec}Spectra of three sample globular clusters (GCs): the young GC W11 and the 
two old metal-poor GCs W12 and W22. Several prominent spectral features are labeled.}

\newpage

\epsfxsize=14cm
\epsfbox{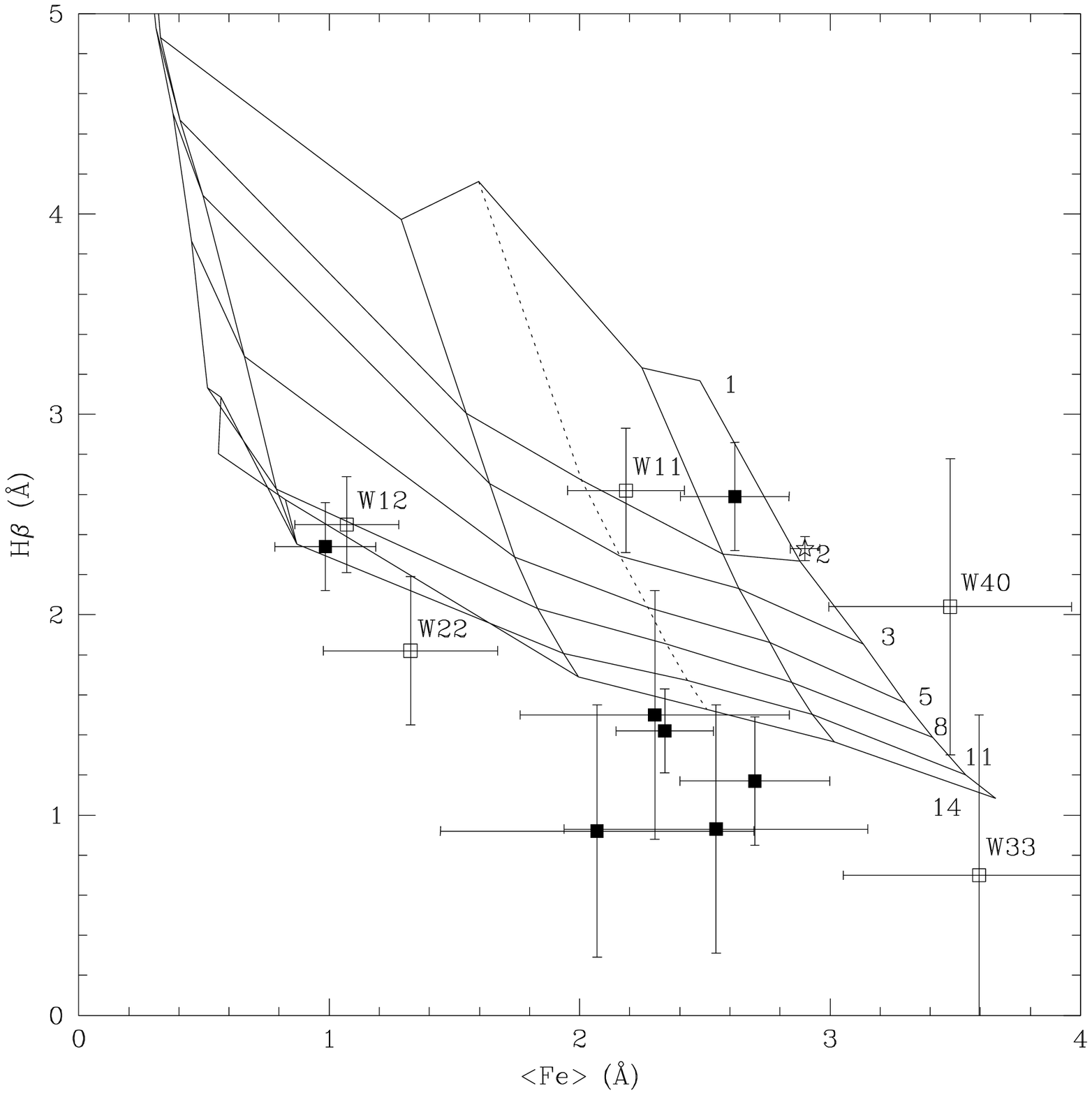}
\figcaption[strader.fig2.eps]{\label{fig:fe_hb}Plot of H$\beta$ vs.\ $<$Fe$>$, with a grid of model isochrones and isometallicity lines by Thomas, Maraston \& Bender
(2003) superposed. From left to right, the isometallicity lines represent [$Z$/H] = $-$2.25, $-$1.35, $-$0.33, 0.00, 0.35, 0.67. For clarity, the solar isometallicity
line is dotted. Ages are indicated at the right of each isochrone. Filled squares are globular clusters from Strader \etal (2003); open squares are from the present
work, and are labeled with their ID. The open star is the r$_{e}/8$ central aperture from Denicolo \etal~(2003). Due to difficulties in measuring $<$Fe$>$, the cluster
W28 is not plotted.}

\newpage

\epsfxsize=14cm
\epsfbox{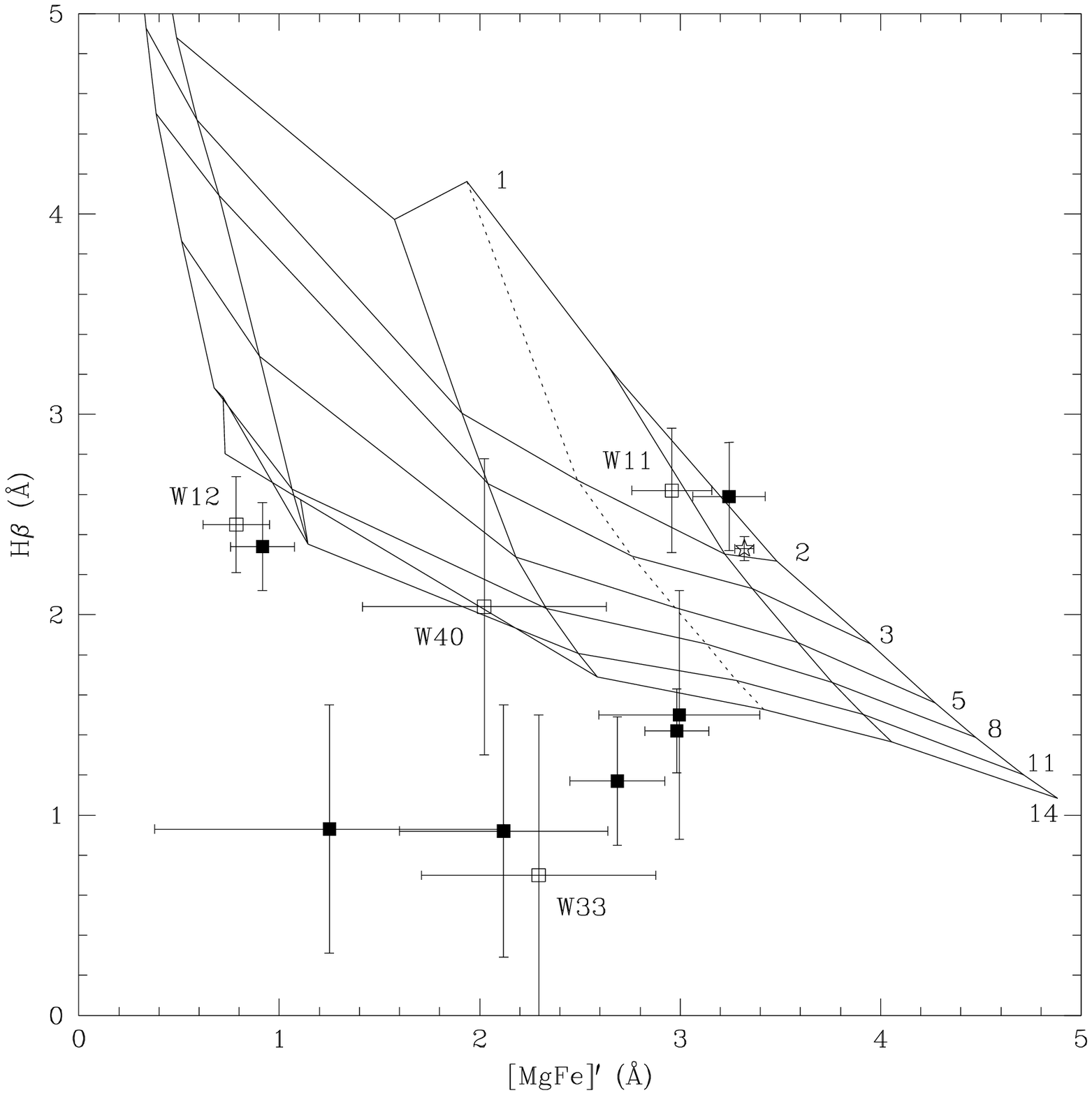}
\figcaption[strader.fig3.eps]{\label{fig:mg2_hb}Plot of H$\beta$ vs.\ [MgFe]$\arcmin$, with a grid of model isochrones and isometallicity lines by Thomas \etal (2003)
superposed.  From left to right, the isometallicity lines represent [$Z$/H] = $-$2.25, $-$1.35, $-$0.33, 0.00, 0.35, 0.67. For clarity, the solar isometallicity line is
dotted. Ages are indicated at the right of each isochrone. Filled squares are globular clusters from Strader \etal (2003); open squares are from the present work, and
are labeled with their ID. The open star is the r$_{e}/8$ central aperture from Denicolo \etal~(2003). Due to difficulties in measuring $<$Fe$>$ and Mg$b$, respectively,
clusters W28 and W12 are not plotted.}

\newpage

\epsfxsize=14cm
\epsfbox{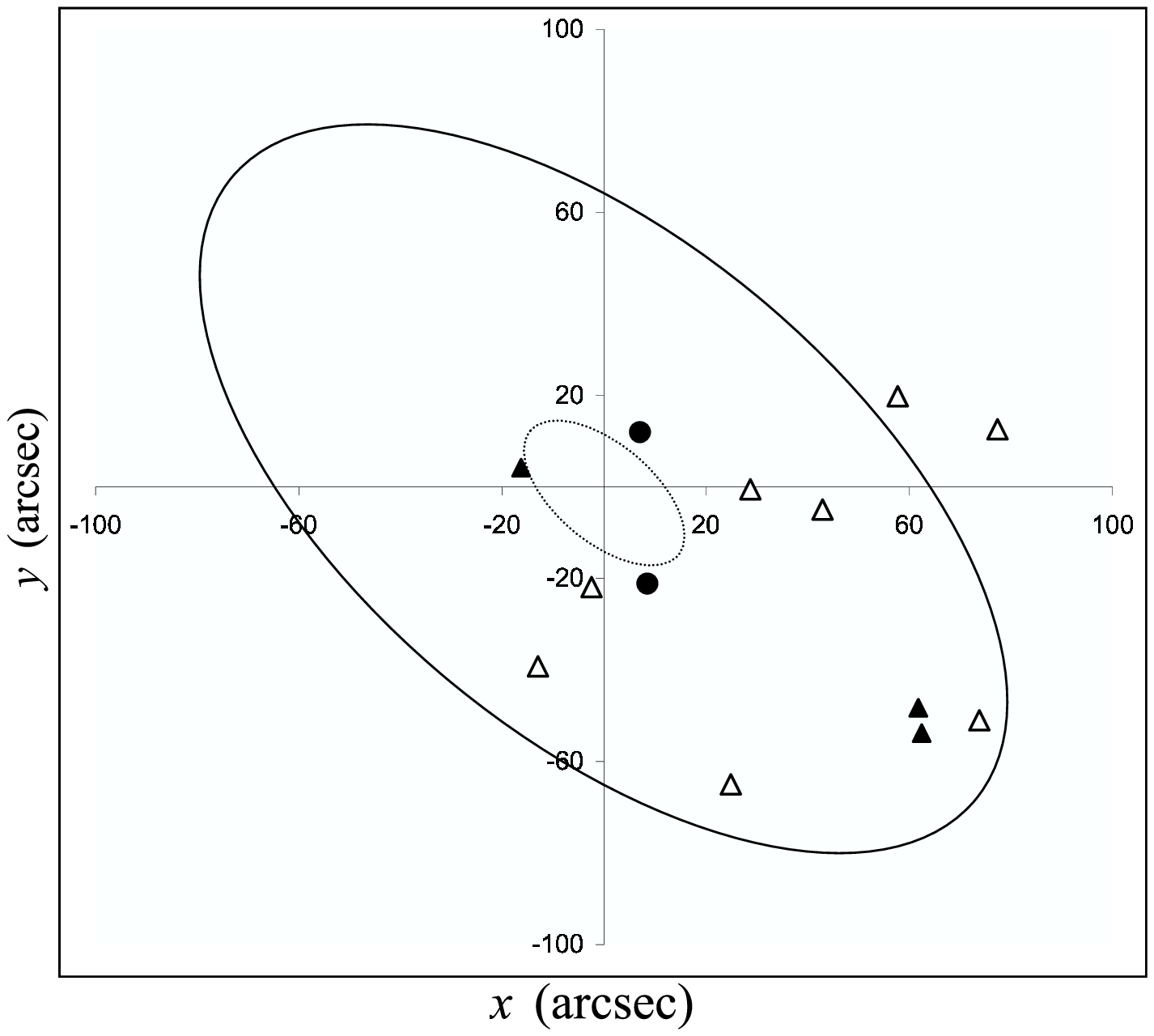}
\figcaption[strader.fig4.eps]{\label{fig:sche}A schematic diagram showing the projected positions of our sample globular 
clusters (GCs) relative to the galaxy center. Open triangles mark old metal-rich GCs, filled triangles mark old metal-poor GCs, and filled circles 
mark young GCs. The inner dotted ellipse traces the $B$ band isophotal effective radius, and the outer solid circle denotes the $K$ band isophotal 
effective radius. Eleven GCs fall within the $K$ band effective radius, with one on the edge of the $B$ band effective radius.}

\newpage

\epsfxsize=14cm
\epsfbox{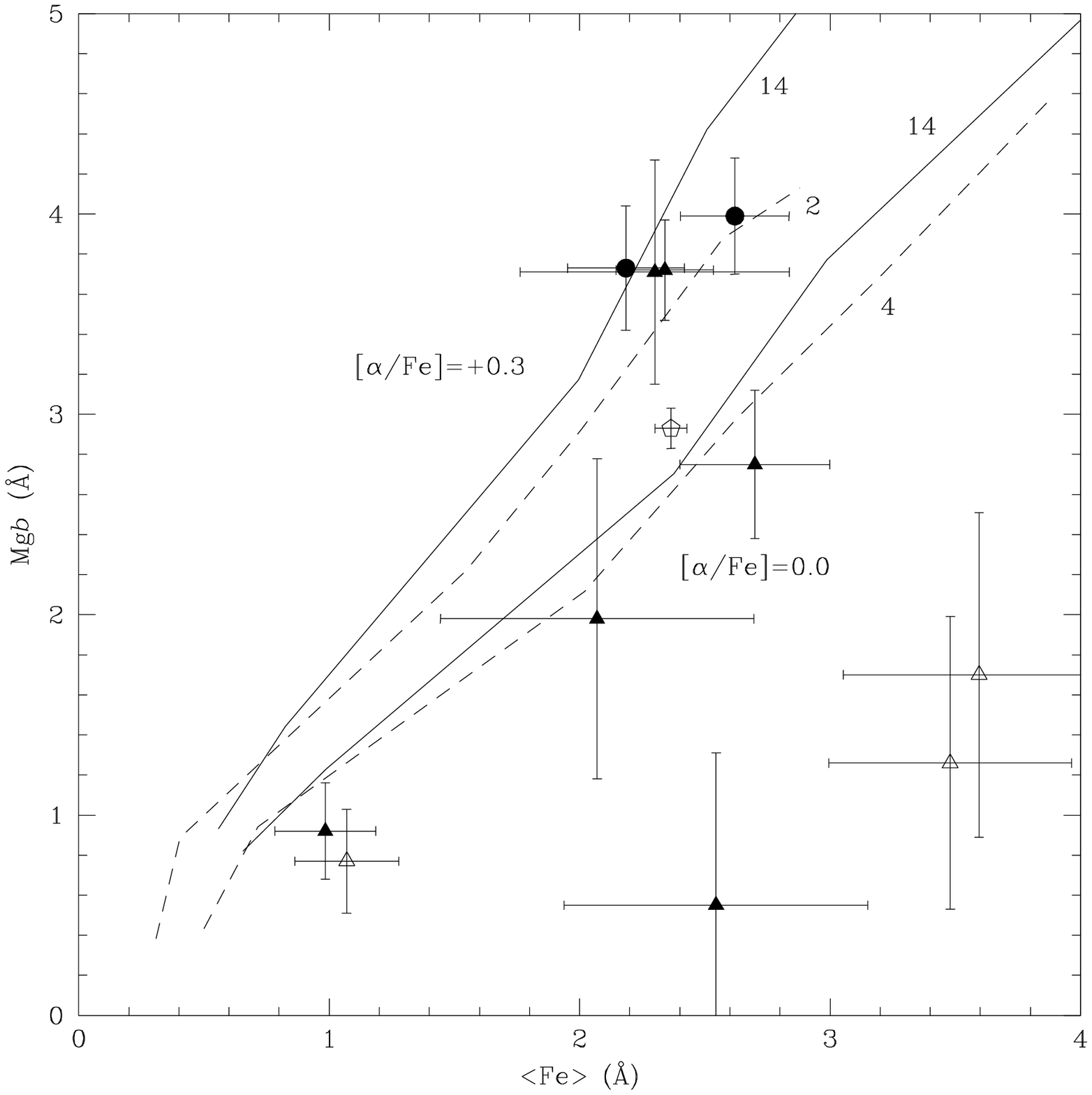}
\figcaption[strader.fig5.eps]{\label{fig:alla}Comparison of Mg$b$ and $<$Fe$>$ for all globular clusters (GCs). 2/4 Gyr (dotted lines) and 14
Gyr (solid lines) isochrones from Thomas \etal (2003), varying in [$\alpha$/Fe], are superimposed. Filled triangles are old GCs from
Strader \etal (2003), open triangles are old GCs from the present work, and large filled circles are young GCs. The large pentagon is the
Galactic open cluster M67. Due to difficulties in measuring $<$Fe$>$ and Mg$b$, respectively, clusters W28 and W12 are not plotted.}

\newpage

\epsfxsize=14cm
\epsfbox{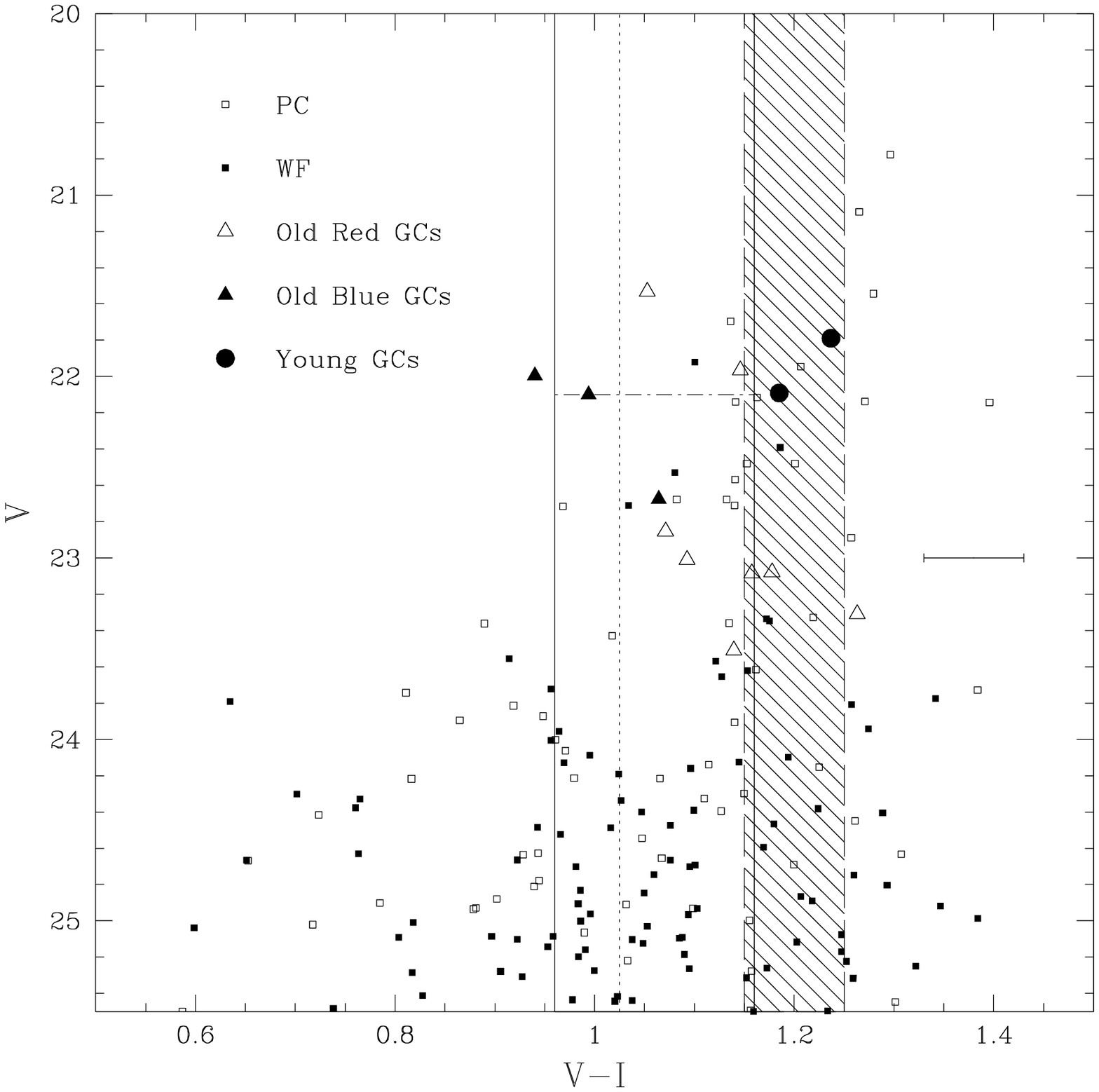}
\figcaption[strader.fig6.eps]{\label{fig:cmdf4}The color-magnitude diagram for all globular clusters (GCs) with $V > 25.5$ and $0.5 < V-I < 1.5$
detected in the HST data by Whitmore \etal (2002; W02). The open squares are objects detected on the PC chip and are closer to the galaxy center.
The filled squares are objects detected on the three WF chips. Open triangles mark the positions of the eight old red GCs in our sample, filled
triangles mark the three old blue GCs, and filled circles mark the young GCs W6 and W11. The dotted line denotes the photometric color cut ($V-I$
= 1.025) adopted by W02 to separate blue and red clusters. The shaded region shows a simulated 1$\sigma$ color distribution of an 1.6 Gyr,
[$Z$/H] = +0.6 population of GCs (using Bruzual \& Charlot 2001 models), and the solid lines define the 1$\sigma$ distribution of old metal-rich
Galactic GCs. The dot-dashed line at $V = 22.1$ shows the magnitude $\omega$ Cen would have at the distance of NGC 3610. A typical $V-I$ error
bar for the clusters in our sample is shown.}

\newpage

\begin{deluxetable}{lccccccc}
\tablewidth{0 pt}
\rotate
\tablecaption{ \label{tab:ident}
  Basic Data for Globular Clusters in NGC 3610\tablenotemark{a}}
\tablehead{ID \tablenotemark{b} & R.A. \tablenotemark{b} & Dec. \tablenotemark{b} & $V$ \tablenotemark{b} & $V - I$ \tablenotemark{b} &
RV \tablenotemark{c} & Proj.~Radius \tablenotemark{b} & S/N \tablenotemark{d} \\
 	&	(hr:min:sec)	&	($^{\circ}$:\arcmin:\arcsec)	&	(mag)	&	(mag)	& (km s$^{-1}$)	& (\arcsec) & }
\startdata

W11 & 11:18:23.79 & 58:47:02.90 & $22.09\pm0.05$ & $1.19\pm0.05$ & $1877\pm28$ & 13.7 & 18.6 \\
W12 & 11:18:24.45 & 58:47:26.87 & $22.10\pm0.05$ & $0.99\pm0.05$ & $1636\pm31$ & 16.8 & 14.8 \\
W22 & 11:18:32.23 & 58:46:15.43 & $22.68\pm0.05$ & $1.06\pm0.05$ & $1692\pm19$ & 78.0 & 24.0 \\
W33 & 11:18:25.69 & 58:46:42.79 & $23.31\pm0.05$ & $1.26\pm0.05$ & $1779\pm37$ & 28.6 & 7.4 \\
W40 & 11:18:23.51 & 58:46:11.92 & $23.51\pm0.05$ & $1.14\pm0.05$ & $1671\pm34$ & 60.7 & 7.6 \\
\tableline
W3  & 11:18:27.96 & 58:47:16.05 & $21.53\pm0.05$ & $1.05\pm0.05$ & $1831\pm20$  & 22.0 & 24.7 \\
W6  & 11:18:28.03 & 58:47:05.11 & $21.79\pm0.05$ & $1.24\pm0.05$ & $1799\pm17$  & 22.7 & 30.4 \\
W9  & 11:18:33.84 & 58:46:53.75 & $21.96\pm0.05$ & $1.15\pm0.05$ & $1825\pm22$ & 69.3 & 36.7 \\
W10 & 11:18:32.95 & 58:46:15.41 & $22.00\pm0.05$ & $0.94\pm0.05$ & $1774\pm32$ & 82.0 & 36.5 \\
W28 & 11:18:24.71 & 58:45:53.33 & $22.85\pm0.05$ & $1.07\pm0.05$ & $1712\pm32$ & 77.9 & 10.4 \\
W30 & 11:18:30.02 & 58:47:28.42 & $23.01\pm0.05$ & $1.09\pm0.05$ & $1724\pm31$ & 41.2 & 13.4 \\
W31 & 11:18:26.45 & 58:46:29.29 & $23.08\pm0.05$ & $1.18\pm0.05$ & $1755\pm21$ & 42.9 & 12.3 \\
W32 & 11:18:32.75 & 58:46:03.85 & $23.09\pm0.05$ & $1.16\pm0.05$ & $1732\pm24$ & 89.2 & 12.9 \\

\enddata
\tablenotetext{a}{The first set of clusters is that studied in the current work; the second set is from Strader \etal (2003).}
\tablenotetext{b}{Taken from Whitmore \etal (2002).}
\tablenotetext{c}{Heliocentric radial velocities determined in this work or in Strader \etal (2003).}
\tablenotetext{d}{Average signal-to-noise ratio per resolution element over the wavelength range 4700--5300 \AA.}

\end{deluxetable} 

\begin{deluxetable}{lrrrrrrrrr}
\tablewidth{0pt}
\rotate
\tabletypesize{\footnotesize}
\tablecaption{Brodie \& Huchra (1990) Individual [Fe/H] Estimates and Composite Values for Newly-Studied Globular Clusters
	\label{tab:metal}}
\tablehead{ ID & $\Delta$ & $\textrm{Mg}_{2}$ & MgH & G Band & CNB & Fe5270 & CNR & H + K & [Fe/H]}
\startdata

W12 & $-1.73\pm0.37$ & $-2.20\pm0.35$ & $-1.70\pm0.51$ & $-1.83\pm0.43$ & $-1.87\pm0.34$ & $-0.39\pm0.65$ & $-0.59\pm0.47$ & $-0.73\pm0.43$ & $-1.45\pm0.29$ \\
W22 & $-1.42\pm0.37$ & $-2.01\pm0.34$ & $-0.91\pm0.49$ & $-1.18\pm0.36$ & $-1.96\pm0.35$ & $-1.90\pm0.62$ & $-1.01\pm0.47$ & $-1.04\pm0.47$ & $-1.46\pm0.20$ \\
W33 & $-1.50\pm0.37$ & $-0.79\pm0.39$ & $-0.56\pm0.59$ & $-1.83\pm0.76$ & $-1.65\pm0.61$ & $-0.38\pm0.75$ & \nodata & \nodata & $-0.88\pm0.48$ \\
W40 & $-0.43\pm0.38$ & $-0.87\pm0.42$ & $-1.07\pm0.62$ & \nodata        & \nodata        & $-0.69\pm0.73$ & $-0.68\pm0.51$ & $-0.91\pm0.61$ & $-0.79\pm0.11$ \\

\enddata
\end{deluxetable}

\begin{deluxetable}{lcccccccccc}
\tablewidth{0pt}
\rotate
\tabletypesize{\footnotesize}
\tablecaption{Lick/IDS Indices\tablenotemark{a}
	\label{tab:indic}}
\tablehead{ID &  H$\beta$ & H$\gamma_{A}$ & H$\delta_{A}$ & CN$_{2}$ & Ca4227 & G4300 & Fe5270 & Fe5335 & $\rm{Mg}_{2}$ & Mg$b$\\
	      &	 (\AA)	  &    (\AA)      & (\AA)         & (mag) & (\AA) & (\AA) & (\AA) & (\AA)  & (mag)         & (\AA)}
\startdata

W11 & $2.62\pm0.31$ & $-4.06\pm0.59$ & $-0.88\pm0.58$ & $0.03\pm0.02$ & $0.59\pm0.33$  & $4.02\pm0.61$ & $2.55\pm0.32$ & $1.82\pm0.34$ & $0.29\pm0.01$ & $3.73\pm0.31$ \\
W12 & $1.82\pm0.37$ & $0.71\pm0.73$  & $-1.61\pm0.61$ & $0.08\pm0.02$ & $-0.03\pm0.37$ & $1.03\pm0.80$ & $1.68\pm0.40$ & $0.97\pm0.57$ & $0.03\pm0.01$ & \nodata \\
W22 & $2.45\pm0.24$ & $0.25\pm0.54$  & $-0.19\pm0.64$ & $0.01\pm0.02$ & $0.07\pm0.34$  & $3.12\pm0.45$ & $0.46\pm0.25$ & $1.68\pm0.33$ & $0.05\pm0.01$ & $0.77\pm0.26$ \\
W33 & $0.70\pm0.80$ & $1.02\pm1.27$  & $-2.27\pm1.17$ & $0.27\pm0.05$ & $2.46\pm0.85$  & $1.72\pm1.73$ & $2.46\pm0.72$ & $4.73\pm0.81$ & $0.15\pm0.02$ & $1.70\pm0.81$ \\
W40 & $2.04\pm0.74$ & $-7.37\pm1.40$ & $1.49\pm1.47$  & $0.11\pm0.04$ & $2.88\pm0.46$  & $12.1\pm1.34$ & $2.95\pm0.68$ & $4.01\pm0.69$ & $0.16\pm0.03$ & $1.26\pm0.73$ \\
\tableline
W3 & $1.17\pm0.32$  & $-3.16\pm0.55$ & $-0.25\pm0.73$ & $0.06\pm0.02$  & $-0.23\pm0.38$  & $3.85\pm0.60$ & $2.53\pm0.37$ & $2.87\pm0.47$ & $0.16\pm0.01$ & $2.75\pm0.37$ \\
W6 & $2.59\pm0.27$  & $-2.90\pm0.33$ & $-1.24\pm0.51$ & $0.13\pm0.01$  & $1.14\pm0.27$   & $5.16\pm0.36$ & $2.66\pm0.28$ & $2.58\pm0.33$ & $0.23\pm0.01$ & $3.99\pm0.29$ \\
W9 & $1.42\pm0.21$  & $-2.22\pm0.39$ & $1.33\pm0.38$  & $0.00\pm0.01$  & $0.82\pm0.22$   & $2.89\pm0.39$ & $2.46\pm0.25$ & $2.22\pm0.30$ & $0.16\pm0.01$ & $3.72\pm0.25$ \\
W10 & $2.34\pm0.22$ & $-0.90\pm0.34$ & $1.09\pm0.41$  & $0.01\pm0.01$  &  $0.24\pm0.19$  & $3.05\pm0.34$ & $0.82\pm0.27$ & $1.15\pm0.30$ & $0.06\pm0.01$ & $0.92\pm0.24$ \\
W28 & $2.28\pm0.81$ & $-0.89\pm1.25$ & $6.16\pm0.97$  & $-0.05\pm0.04$ &   $2.61\pm0.70$ & $5.53\pm1.21$ & $3.07\pm0.79$ & \nodata       & $0.06\pm0.02$ & $2.13\pm0.84$ \\
W30 & $1.50\pm0.62$ & $-3.99\pm1.06$ & $0.56\pm0.94$  & $0.11\pm0.03$  &   $0.47\pm0.27$ & $5.43\pm0.75$ & $2.57\pm0.67$ & $2.03\pm0.84$ & $0.18\pm0.02$ & $3.71\pm0.56$ \\
W31 & $0.92\pm0.63$ & $-3.82\pm0.90$ & $5.82\pm1.09$  & $0.01\pm0.02$  &  $-0.18\pm0.73$ & $5.95\pm0.85$ & $2.52\pm0.79$ & $1.62\pm0.97$ & $0.11\pm0.02$ & $1.98\pm0.80$ \\
W32 & $0.93\pm0.62$ & $-3.14\pm1.04$ & $-2.14\pm0.61$ & $-0.01\pm0.03$ &  $0.55\pm0.55$  & $2.29\pm1.13$ & $3.22\pm0.64$ & $1.87\pm1.03$ & $0.10\pm0.02$ & $0.55\pm0.76$ \\

\enddata

\tablenotetext{a}{The first set of clusters is that studied in the current work; the second set is from Strader \etal (2003).}

\end{deluxetable}

\end{document}